\documentclass[aps,pre,twocolumn,groupedaddress]{revtex4}
\usepackage{graphicx}
\usepackage{dcolumn}
\usepackage{bm}
\usepackage[centertags]{amsmath}
\usepackage{amsfonts}
\usepackage{amssymb}
\usepackage{amsthm}
\usepackage{newlfont}
\usepackage{color}

\begin{document}

\title{Construction of coarse-grained models by reproducing equilibrium probability density function}
\author{Shijing Lu}
\affiliation{Institute of Modern Physics, Zhejiang University, Hangzhou 310027, China}

\author{Xin Zhou
\footnote{To whom correspondence should be addressed. Electronic Mail: xzhou@ucas.ac.cn}
}

\affiliation{School of Physics, University of Chinese Academy of Sciences, Beijing 100049}

 
\date{\today}

\begin{abstract}
The present work proposes a novel methodology for constructing coarse-grained (CG) models, which aims at minimizing the difference between the CG model and its original system. The difference is defined as a functional of the ratio of equilibrium conformational probability densities to the original one, then is further expanded by equilibrium averages of a set of sufficient and independent physical quantities as basis functions. An orthonormalization strategy is adopted to get the independent basis functions from sufficiently preselected interesting physical quantities of the system. The probability density matching coarse-graining (PMCG) scheme effectively takes into account the overall characteristics of the original systems to form CG models, and 
 it is a natural improvement of the usual CG scheme wherein equilibrium averages of many physical quantities are intuit chosen to reproduce without considering correlations among these quantities.  
We verify the general PMCG framework in constructing a one-site CG water model from TIP3P model. Both structure of liquids and pressure are found to be well reproduced at the same time in the one-site CG water model. 

\end{abstract}


\maketitle{}

\section{INTRODUCTION} 
\label{sec:introduction}
Two branches of computational physics, namely macro-scale continuous fluid dynamics and micro-scale molecular dynamics simulations, never stop efforts in extending their power to wider ranges, and to cover the gap between their accessed scales. 
For example, molecular dynamics simulation has been successfully applied to model mesoscopic convection 
and by the same token fluid dynamics has been applied to simulate nano-scale convection behavior. 
However, even though the proliferation of multiprocessor computers and parallel computation techniques have inspired the ambition of computer scientists, molecular dynamics simulation is still an torment for most large scale systems like macromolecules' self-assembling problem whose typical size is in micrometer 
and whose typical time is in microsecond or longer~\cite{klein2008}. In this circumstance coarse-grained (CG) method, as a promising way, and in some situations the only way~\cite{klein2008}, of bringing molecular dynamics simulation into large applications on multi-$\mu$s time scale or multi-$\mu$m length scale has therefore been discussed extensively recent years~\cite{Praprotnik2008, Stewart2006, Sergei2005}. Different from all-atom (AA) molecular dynamics, CG method usually subsumed high frequency intramolecular vibrations into coarse grained sites. The lower resolution model is then simulated in a carefully designed effective force field. This procedure has been proved available for many problems, such as protein dynamics~\cite{Klepeis2009120}, protein-membrane interaction simulation~\cite{Lindahl2008425} as well as protein folding~\cite{Clementi200810}. 

Usually, the construction of a CG model includes three steps. The first one is to select the mapping from AA coordinates to CG coordinates, simply denoted as $x=x(q)$. Here $q$ and $x$ represent the high-dimensional conformation vectors in AA and CG models, respectively. For example, while $q$ is the position vector of all hydrogen and oxygen atoms of waters, $x$ can be chosen as the position vector of the center of mass of water molecules, thus $x(q)$ is a linear function of $q$. Another example of CG coordinates $x$ is the spatial distribution of particle density $\rho(\mathbf{R})$, or the distribution of a vector, such as in liquid crystals, $\mathbf{n}(\mathbf R)$, while $q$ is still the atomic coordinates of these molecules. Thus CG model could be a field-based continuous model even the original model is particle-based;   
The second step is to presume a formula of effective interaction in the CG space, $x$. The effective interaction can be generally written as $U(x; \lambda)$ with a set of parameters, $\lambda$; 
The final step of constructing CG models is to determine the parameters of the effective CG interactive potential by minimizing a pre-defined distance between CG model and the referenced AA system. 
While the first two steps are usually based on afore experience and knowledge of the studied systems,  
most current coarse-graining techniques focus on the optimization of CG potential in the parameter space, based on different measurement of inter-model distance~\cite{Noid2008,Scott2008}. 
 
Traditionally, coarse-graining methods measure the distance between CG models and the originals based on equilibrium averages of some intuit chosen physical quantities, such as the radius distribution function (RDF) of CG sites~\cite{Mueller-Plathe2002}, 
\begin{eqnarray}
D_{trad}= \int [\langle {\hat g}(r; x) \rangle_{cg} - \langle {\hat g}(r; x) \rangle_{aa}]^{2} w(r) d r, 
\label{distance-trad}
\end{eqnarray}
where $\langle {\hat g}(r; x) \rangle_{cg/aa} \equiv g_{cg/aa}(r)$ is the usual RDF of CG sites in the CG/AA model, which is the ensemble average of microscopic RDF, ${\hat g}(r; x)$.  $w(r)$ is an optional weight, which might affect measurement of the distance thus the optimized parameters. Usually ones set the weight function based on experience or some physical consideration.  
The traditional CG method has already been widely applied in various systems~\cite{SherwoodBS2008,DelleSiteAAK2002,ZhouADK2005,ZhouADK2005b}. 
Two known algorithms, iterative inverse Boltzmann~\cite{ib:jcc2002} and reverse Monte Carlo~\cite{rmc:pre1995} are applied in the traditional CG techniques to optimize CG force field. The iterative inverse Boltzmann method, introduced by Reith \textit{et al.}~\cite{ib:jcc2002} iteratively adjust the CG inter-particle interactions $u(r)$ based on  the iteration relation, $g_{cg}(r)=\exp[-\beta u(r)]$, between the pairwise potential function $u(r)$ and the equilibrium averaged RDF $g_{cg}(r)$ of CG sites, until the $g_{cg}(r)$ is approach to the target $g_{aa}(r)$. Although the relation is exact only in the low density limit, the iterative inverse Boltzmann can be valid for dense systems like Lennard-Jones liquid~\cite{ib:jcc2002}, TIP3P water~\cite{wang2009} and polymer systems~\cite{ib:jcc2002, Faller20043869}. Similar to the iterative inverse Boltzmann method, the reverse Monte Carlo (RMC) method~\cite{rmc:pre1995}, also consists of iterative adjustment of interactions, only that modification to potential function in each RMC iteration cycle is determined based on 
the estimated derivatives of the equilibrium averages of the chosen physical properties in the parameter space, {\emph i.e.}, it minimizes the distance such as eq.(\ref{distance-trad}) by some derivative-based optimization methods, such as steepest descent method or conjugate gradient method. RMC has been used to determine ion-ion interaction in aqueous NaCl solutions~\cite{rmc:pre1995}, to study diffusive dynamics of liquid water~\cite{matysiak024503}, to model behavior of mesoscopic lipids and lipids assembly in water~\cite{Lyubartsev2005} and to study bilayer membrane~\cite{murtola075101}.

However, the optimized CG model based on the traditional CG method may be dependent on the selection of physical quantities. Usually, ones select some interesting physical variables of the system, such as RDF in identical particles systems~\cite{Mueller-Plathe2002}, distribution of soft torsion angles in chain molecules~\cite{DelleSiteAAK2002}, etc. It is not clear how another variables are reproduced in the formed CG models. The weights of the physical variables, {\textit e.g.} $w(r)$ in eq.(\ref{distance-trad}), may also affect the optimized CG potential, so it is also important to know which $w(r)$ is reasonable. In addition, in the traditional CG methods, the fitted properties are macroscopic (ensemble) average values. It is unknown how the formed CG model reproduces the microscopic distribution of the corresponding physical variables. 
 To improve the problems, in a previous work~\cite{ZhouJZR2008}, we presented 
  an alternative CG method matching the overall free energy surface of the AA model in the CG conformational space. The free energy surface is defined as,   
\begin{equation}
  \label{free-energy}
A(x) = -k_{B} T \ln \int \exp[-\beta V(q)] \delta(x-x(q)) dq, 
\end{equation}
where $k_{B}$ is the Boltzmann's constant and $T$ is temperature, $\beta=1/k_{B} T$, $V(q)$ is the potential energy surface of the AA system, and $\delta()$ is the Dirac-$\delta$ function. 
Thus the distance between the CG model and AA system could be defined as 
$D_{fe}=\langle [\Delta U(x)]^{2} \rangle_{P(x)}$,
where $\Delta U(x) \equiv U(x;u_{\gamma}) - A(x)$, and $U(x;u_{\gamma})$ is the effective CG potential, $P(x)$ is an optional probability density, such as the equilibrium distribution of AA system ($P_{aa}(x) \propto e^{-\beta A(x)}$), or that of CG model ($P_{cg}(x) \propto e^{-\beta U(x;u_{\gamma})}$), or other related distribution.  
By calculating values of the free energy $A(x)$ at many sampled CG conformations, such as $\{x^{i}\}, i=1, \cdots, M$, based on a jump-in-sample algorithm~\cite{ZhouJZR2008}, the distance was estimated as $D_{fe} = \frac{1}{M} \sum_{i} [\Delta U(x^{i})]^{2}$, then an effective CG force field of tetrahedral molecular liquids was optimized~\cite{ZhouJZR2008}. 
The free-energy matching method is expected to better take into account the overall characteristics of the AA system in comparison with the traditional CG methods~\cite{ZhouJZR2008}, however more computational costs are usually required in the free-energy matching method than the latter. 
Replacing to the very time-consuming calculation of $\{A(x^{i})\}$, it is possible to calculate the gradients of free energies, $\frac{\partial A}{\partial x}|_{x=x^{i}}$, for example based on the blue-moon ensemble simulations~\cite{CiccottiKV2002}, to define the distance between CG and AA models, $D_{feg} = \frac{1}{M} \sum_{i} |\frac{\partial}{\partial x} \Delta U(x^{i})|^{2}$, and to optimize CG force field with less computational cost. 
By matching the total AA force on CG sites, Voth's group presented a CG method, MSCG~\cite{Ercolessi1994,Izvekov2004}, to form CG molecular models from AA models in various of systems by using linear CG mapping function $x=x(q)$ (see \cite{Noid2008} and references therein). 
For the linear mapping function, the gradients of free energy matching are actually same as the force matching method. 
 
In the work we present a new variant of the free-energy matching method in construction of CG model, named as probability density matching CG method (PMCG), which reproduces the equilibrium probability density function of CG model, $P_{cg}(x) \propto e^{-\beta U(x)}$, to that of AA system, $P_{aa}(x) \propto e^{-\beta A(x)}$ . A functional of the ratio of the two probability density functions, ${\cal F}[e^{-\beta \Delta U(x)}]$, is found to accurately measures the distance between the two models, since it provides an upper limit of the error of reproducing the AA equilibrium average of any conformational function in the CG model. The functional can be expanded based on a set of complete and linearly independent conformational functions. Thus, PMCG is a natural improvement of the traditional CG method by orthonormalization of selected physical quantities in the latter, so PMCG has almost same computational cost as the latter in practice. On the other hand, the functional of probability density functions, ${\cal F}[e^{-\beta \Delta U(x)}]$,  
takes into account the overall difference between $A(x)$ and $U(x)$ in whole the CG conformational space, $x$, as the free-energy-based CG methods (by matching free energy or its gradient or force).  
Therefore,  
PMCG shows to have the advantages of both the traditional CG method and the free-energy-based CG methods. 
  
This article is organized as follows, in Sec.~\ref{sec:theory}, we first introduce the theories on definition of distance  between CG model and its original, and then give the complete scheme of PMCG. 
 In Sec.~\ref{sec:experiment} we demonstrate the application of PMCG in liquid water to form a one-site CG water model  which not only well reproduces the structure of AA liquid (radial distribution function) but its pressure at the same time. 
 
\section{THEORY AND METHODS} 
\label{sec:theory} 

\subsection{Distance between models}
While groups of atoms in a AA system are mapped as CG sites, {\it i.e.}, $x=x(q)$, the interaction among CG sites, $U(x;u_{\gamma})$, need to be optimized in the parameter $u_{\gamma}$ space to minimize the difference between CG model and the AA system. 
It possibly requires $U(x;u_{\gamma})$ matches the free energy surface $A(x)$ in important (and interesting) conformational regions of $x$. 
We define the ratio of two probability density functions,
\begin{eqnarray}
\Omega_{aa \rightarrow cg}(x) \equiv \frac{P_{cg}(x)}{P_{aa}(x)},
\end{eqnarray}
where $P_{cg/aa}(x)$ is the equilibrium probability density function of the CG/AA model in the $x$ space, for example, $P_{cg}(x) \propto e^{-\beta U(x;u_{\gamma})}$ and $P_{aa}(x) \propto e^{-\beta A(x)}$ in canonical ensemble. 
$\Omega_{aa \rightarrow cg}(x)$ characterizes the difference of the CG model from the AA models. Considering the fact 
\begin{eqnarray}
\langle \Omega_{aa \rightarrow cg}(x) B(x) \rangle_{aa} = \langle B(x) \rangle_{cg}, 
\end{eqnarray}
for any conformational function $B(x)$,  
the ratio function can be expanded as a series of complete basis functions $B^{\mu}(x)$~\cite{GongZ2009,GongZ2010}, 
\begin{equation}
  \label{fee:expansion}
  \Omega_{aa \rightarrow cg}(x) = 1 + \sum_{\mu\nu} G_{\mu\nu}(aa) \; \langle \delta_{aa}B^{\mu}(x) \rangle_{cg} \; \delta_{aa} B^{\nu}(x). 
\end{equation}
Here $\langle \cdots \rangle_{aa/cg}$ denotes the ensemble average in the AA/CG model,
 $\delta_{aa} B^{\mu}(x) \equiv B^{\mu}(x)- \langle B^{\mu}(x) \rangle_{aa}$,  
 $G_{\mu\nu}(aa)$ is the inverse covariance matrix of the basis functions which obey the equality,  $\sum_{\mu} G_{\lambda\mu}(aa) G^{\mu\nu}(aa) = \delta_{\lambda}^{\nu}$ while 
 $G^{\mu\nu}(aa) \equiv \langle \delta_{aa} B^{\mu}(x) \; \delta_{aa} B^{\nu}(x) \rangle_{aa}$ is the covariance matrix of the basis functions in the AA model. 
 The fact $\langle \Omega(x) \rangle_{aa} =1$ is explicitly wrote out in eq.(\ref{fee:expansion}) by selecting the first basis function $B^{0}(x) \equiv 1$. 
We can define the difference from the AA model to CG model as the equilibrium fluctuation of the ratio function in the AA model, 
\begin{eqnarray}
  \label{eq:s2}
  s^2_{aa \rightarrow cg} &\equiv& \langle \Omega^{2}_{aa \rightarrow cg}(x) \rangle_{aa} -1 \nonumber \\
  &=& \sum_{\mu\nu} G_{\mu\nu}(aa) \; \langle \delta_{aa} B^{\mu} \rangle_{cg} \; \langle \delta_{aa} B^{\nu} \rangle_{cg} \nonumber \\
  &=& \sum_{\mu\nu} G_{\mu\nu} [\langle B^{\mu}\rangle_{cg} - \langle B^{\mu} \rangle_{aa}][\langle B^{\nu}\rangle_{cg} - \langle B^{\nu} \rangle_{aa}], 
\end{eqnarray}
because $s_{aa \rightarrow cg}$ gives the upper limit of the relative error of the ensemble means of any variable $B(x)$, 
\begin{eqnarray}
\epsilon &=& |\langle B(x) \rangle_{cg} - \langle B(x) \rangle_{aa}| = 
|\langle (\Omega-1) \delta_{aa}B \rangle_{aa} | \nonumber \\
&\leq& s_{aa \rightarrow cg} \sigma^{\mu}(aa), 
\label{error}
\end{eqnarray} 
where $\sigma^{\mu}(aa) = \sqrt{ \langle (\delta_{aa} B^{\mu})^2 \rangle_{aa} }$ is the fluctuation of $B^{\mu}(x)$ in the AA system. 
It is worth to mention that $s_{aa \rightarrow cg}$ is not a normal distance, since it is not symmetric about $aa$ and $cg$, {\it i.e.}, $s_{cg \rightarrow aa} \neq s_{aa \rightarrow cg}$. It is not difficult to define a symmetric distance, such as, $D^{2}(aa,cg) = 4 (\langle \Omega_{{\bar P} \rightarrow cg}^{2} \rangle_{\bar P} -1)$, where ${\bar P}(x) = [P_{aa}(x) + P_{cg}(x)]/2$ is the average probability density and $\Omega_{{\bar P} \rightarrow cg}(x) \equiv \frac{P_{cg}(x)}{\bar P(x)}$.  
For theoretical view points, it might be more robust to use the symmetric distance $D^{2}(aa,cg)$ in the expansion of the ratio of probability density functions. 
The minimization of $D^{2}(aa,cg)$ is similar to that of $s^{2}_{aa \rightarrow cg}$ except needing slightly more computational cost in comparison with the latter. 
In the current paper, we use the $s^{2}_{aa \rightarrow cg}$ to illustrate the optimization of CG force field, it is direct to extend the method to use $D^{2}(aa,cg)$.  
Actually, the only difference among $s^{2}_{aa \rightarrow cg}$, $s^{2}_{cg \rightarrow aa}$ and $D^{2}(aa,cg)$ is that the inverse covariance matrix $G_{\mu\nu}$ are estimated in different equilibrium conformational samples, such as 
$s^{2}_{aa \rightarrow cg} = \sum G_{\mu\nu}(aa) b^{\mu} b^{\nu}$, $s^{2}_{cg \rightarrow aa} = \sum G_{\mu\nu}(cg) b^{\mu} b^{\nu}$ and $D^{2}(aa,cg)=D^{2}(cg,aa) = \sum G_{\mu\nu}({\bar P}) b^{\mu} b^{\nu}$. Here $b^{\mu} \equiv \langle B^{\mu} \rangle_{cg} - \langle B^{\mu} \rangle_{aa}$.  

If using the microscopic RDF, ${\hat g}(r; x)$, as the basis function in eq.(\ref{eq:s2}) and comparing it with the tradition distance defined at eq.(\ref{distance-trad}), 
it is very clear that the current method considers the cross correlation among the basis function, and use the inverse correlation matrix as the weight. Obvious, if ${\hat g}(r; x)$ at different $r$ are orthogonal, {\emph i.e.} $\langle \delta_{aa} {\hat g}(r; x) \delta_{aa} {\hat g}(r^{\prime}; x) \rangle_{aa} = \sigma^{2}(r) \delta(r-r^{\prime})$, the current method is back to the traditional one with the particular selection $w(r) = \sigma^{-2}(r)$ in eq.(\ref{distance-trad}). 
Therefore, in the current method, we can select many basis functions, the inverse correlation matrix automatically remove linear dependent basis functions and set suitable weights for the independent basis functions.  
 
In principle, if the basis set $\{B^{\mu}(x)\}$ is complete, eq.(\ref{fee:expansion}) is exact and independent on the applied CG potential $U(x)$, thus we have an analyzed expansion of whole the free energy surface in (any) high-dimensional $x$ space, 
\begin{eqnarray}
A(x)&=&U(x) + k_{B} T \ln [ 1 + \sum G_{\mu\nu}(aa) b^{\mu} \; \delta_{aa}B^{\nu}(x) ] \nonumber \\
&=&U(x) - k_{B} T \ln [1 - \sum G_{\mu\nu}(cg) b^{\mu} \; \delta_{cg}B^{\nu}(x) ]. 
\label{free-energy-expansion}
\end{eqnarray}
However, in practice, we only can use a finite-size basis set (and finite-size AA/CG equilibrium samples to get linear independent basis functions) in the expansion, thus the obtained $\Omega(x)$ (and $A(x)$) may have significant errors at somewhere of the $x$ space unless the dimension of $x$ is small and/or $U(x)$ very closes to $A(x)$. 
Although the expansion can not be expected to be accurate in every where of $x$ space, 
$s^{2}$ ( or $D^{2}$) is expected to estimate very well due to averaging process in CG samples in eq.(\ref{eq:s2}), if many (but not very large number of) basis functions are applied in the expansion. 
In practice, we may chose interesting physical quantities of system as much as possible as basis functions to capture the difference between AA and CG models. 
Since some of basis functions possibly linearly dependent on others, the covariant matrix $G^{\mu\nu}$ will not have a full rank. In this situation, we use Gram-Shmit process to form orthogonal basis functions. If both CG and AA models are specific, in the Gram-Shmit process, we include more and more orthogonal basis functions according to their contributions to $s^{2}$ from large to small until no new orthogonal basis function with significant contribution to $s^{2}$ is obtained~\cite{GongZ2009,GongZ2010}.  
As more and more orthogonal basis functions are included, 
the difference between CG and AA models is gradually approached, $s^{2}$ may be saturated~\cite{GongZ2009}, new increased basis functions are not long independent on the existed ones.  
Due to the fact that a finite-size conformational sample in AA system is used to estimate $G_{\mu\nu}(aa)$, the number of independent basis functions is not more than the size of the sample. 
Actually, unless the difference between AA and CG models is very huge, and/or the selected basis functions are completely unable to characterize the difference, the number of required basis functions to measure $s^{2}$ is far smaller than the size of sample. 
In extreme cases, it is possible a few carefully chosen physical quantities might have already been sufficient to estimate $s^{2}$, another basis functions are either linearly dependent on the quantities or very small in contributing $s^{2}$. 
More detailed discussion about the completeness of basis function in estimate of $s^{2}$ is discussed in the reference~\cite{GongZ2009,GongZ2010} where we applied the same expansion in detecting meta-stable structure and enhanced sampling conformational space of complex systems. 
 
In the paper, instead of directly using Gram-Shmit orthogonalization to form independent basis functions~\cite{GongZ2009},  
we calculate the inverse covariance matrix $G^{\mu \nu}$. Because $G^{\mu \nu}$ is positive definite and symmetric, it has all positive eigenvalues and can be decomposed to a form $G^{\mu \nu} = \mathbf{V}^T \mathbf{D} \mathbf{V}$ where $\mathbf{V}$ is a unitary matrix whose rows are orthonormal, $\mathbf{D}$ is diagonal matrix whose diagonal elements are eigenvalues of $G^{\mu \nu}$. In practice, some of the basis functions can be linearly dependent on the others, therefore the covariance matrix $G^{\mu \nu}$ has some zero eigenvalues and in that case $\mathbf{D}$ irreversible. However if we choose row vectors of $\mathbf{V}$ as basis functions and desert those with zero length, we are able to build a non-degenerated functional space. Noticing that deserting eigenvectors are the same as truncating eigenvalues. We define the inverse matrix of $\mathbf{D}$ as:
\begin{equation}
  \label{eq:invD}
  [\mathbf{D}^{-1}]_{ii} = \left\{
    \begin{array}{rl}
      1 / \mathbf{D}_{ii}  & \text{if } \mathbf{D}_{ii} \geq \varepsilon_D \\
      0 & \text{if } \mathbf{D}_{ii} < \varepsilon_D 
    \end{array} \right.
\end{equation}
Here $\varepsilon_D$ is a threshold for truncating eigenvalues ( in our case, $\varepsilon_D$ is set to $0.001$). Then, the inverse covariant matrix writes as $G_{\mu\nu} = \mathbf{V}^T \mathbf{D}^{-1} \mathbf{V}$. An alternative method to set $G_{\mu\nu} = \mathbf{V}^T [\mathbf{D} + \varepsilon_{D} \mathbf{I}]^{-1} \mathbf{V}$, where ${\mathbf I}$ is the unit matrix. 

\subsection{Optimization of CG models}

Now the difference between two models (actually, two finite-size equilibrium conformational samples) is calculable, the only thing left for constructing CG models is to minimize $s^2$ by consecutively adjusting the parameters $u_{\gamma}$ of CG potential $U(x,u_{\gamma})$. 
Very generally, we can suppose $U(x,u_{\gamma}) = \sum_{\gamma} u_{\gamma} f^{\gamma}(x)$, thus 
the gradient of $s^{2}$ is given as, 
\begin{equation}
\label{fee:gradient}
 {\partial s^2}/{\partial u_{\gamma}}
 =  - 2 \beta \sum_{\mu} \langle {\hat B}^{\mu}(x) \rangle_{cg} 
 [\langle {\hat B}^{\mu}(x) f^{\gamma}(x) \rangle_{cg} - \langle {\hat B}^{\mu}(x) \rangle_{cg} \langle f^{\gamma}(x) \rangle_{cg}]. 
\end{equation} 
Here we already supposed ${\hat B}^{\mu}(x)$ is the orthogo-normalized basis functions after the orthogonalization and normalization process, thus 
$\langle {\hat B}^{\mu}(x) \rangle_{aa} = 0$, and $\langle {\hat B}^{\mu}(x) {\hat B}^{\nu}(x) \rangle_{aa} = \delta_{\mu\nu}$. 
Therefore, lots of local optimization methods, such as the steepest descent method, the conjugate gradient method, can be used to minimize $s^{2}$. 
Iteration method directly based on eq.(\ref{fee:expansion}) is also possible to update CG potential by $u_{\gamma} \leftarrow u_{\gamma} + \Delta u_{\gamma}$, where $\Delta u_{\gamma}$ satisfies 
\begin{equation}
  \label{fee:iteration}
  -\beta \sum_{\gamma}f^{\gamma}(x) \Delta u_{\gamma}  = \ln[\Omega(x)]. 
\end{equation} 
In practice, $\Delta u_{\gamma}$ can be solved from the linear equation,    
\begin{equation}
  \label{fee:iteration2}
  \sum_{\gamma} \langle f^{\lambda}(x)f^{\gamma}(x) \rangle_{X} \Delta u_{\gamma}  
= - k_{B} T \langle f^{\lambda}(x) \ln[{\tilde \Omega}(x)] \rangle_{X}. 
\end{equation}
The angular bracket with subscript $X$ means the ensemble average in an optional conformational sample, such as the equilibrium sample in AA or CG model, or any hybrid sample which consisted of equilibrium conformations in both AA and CG models. 
As we mentioned, due to the incomplete basis functions in eq.(\ref{fee:expansion}), it is possible the estimated 
$\Omega(x)$ is very closed to zero or even be negative at some $x$, although it should be positive anywhere in principle. 
It is the reason we replace $\Omega(x)$ by ${\tilde \Omega}(x)$ in eq.(\ref{fee:iteration2}), where ${\tilde \Omega}(x) = \Omega(x)$, if $\Omega(x)$ is larger than a presumed small positive value $\omega_{\epsilon}$, but ${\tilde \Omega}(x) = \omega_{\epsilon}$ if $\Omega(x) < \omega_{\epsilon}$. Here $\omega_{\epsilon}$ is applied to make sure the left side of eq.(\ref{fee:iteration}) be a real number.  In this paper, we set $\omega_{\epsilon}=\mathrm{0.001}$.
In addition, we may use the approximation $\ln [\Omega(x)] \approx \Omega(x) -1$ 
 in eq.(\ref{fee:iteration2}) to get a simpler linear equation,
\begin{equation}
  \label{fee:iteration3}
  \sum_{\gamma} \langle f^{\lambda}(x) f^{\gamma}(x) \rangle_{X} \Delta u_{\gamma}  
= - k_{B} T \sum_{\mu} \langle f^{\lambda}(x) {\hat A}^{\mu}(x)\rangle_{X} \langle {\hat A}^{\mu}(x) \rangle_{cg},  
\end{equation}
for updating the CG potential. Although the approximation is correct only while $\Omega(x)$ closes to $1$, eq.(\ref{fee:iteration3}) is still sufficient as an iteration formula. 
 
As an example, if the effective potential of CG model is pairwise additive and central, $U(x) = \sum_{i<j} u(r_{ij})$, where 
$r_{ij} = |\mathbf{R}_{i} - \mathbf{R}_{j}|$ represents the distance between the $i$th and $j$th CG site. 
We rewrite $U(x)$ as,
\begin{eqnarray}
U(x) \approx \sum_{\gamma} u_{\gamma} f^{\gamma}(x),
\end{eqnarray}
where 
$\{u_{\gamma}\}$, values of pair interaction $u(r)$ at discrete distances $\{r^{\gamma}\}$, are the parameters in minimizing $s^{2}$. $f^{\gamma}(x)= \sum_{i<j} {\tilde \delta}(r_{ij} - r^{\gamma})$. 
 $\tilde{\delta}(r)$ is a smooth version of the double-side step function that $\tilde{\delta}(r)=1$ while  
 $-\Delta r/2 < r < \Delta r/2$, otherwise is zero. 
 Different formula of $\tilde{\delta}(r)$ does not significantly change interesting results.
Thus $f^{\gamma}(x)$ is actually the number of the CG-particle pairs with the pair distance 
in the interval $(r^{\gamma}-\Delta r/2, r^{\gamma}+\Delta r/2)$.
  
Start from eq.(\ref{fee:gradient}) or eq.(\ref{fee:iteration2}) or eq.(\ref{fee:iteration3}), an effective CG potential can be obtained by following algorithms:
\begin{enumerate}
\item Choose an initial guess of potential function $u^{(0)}_{\gamma}$. For example, we might use logarithmic RDF as a  start point in pairwise additive models,  
\begin{eqnarray}
\label{eq:initialGuess}
  u^{(0)}_{\gamma} = -k_{B} T \ln g(r^{\gamma}), 
\end{eqnarray}  
where $g(r^{\gamma})$ is RDF at the pair distance $r^{\gamma}$ of CG sites in the AA system. 
\item Run a MD simulation under the potential with the parameters formed in the $i$th iteration, $\{u^{(i)}_{\gamma}\}$, calculate values of $\{f^{\gamma}(x)\}$ and the orthonormalized basis functions $\{{\hat B}^{\mu}(x)\}$ in the sampled conformations (if using the sample of AA system, the orthonormalization only need to do once at the $i=0$ step), and estimate $s^2$ (and its gradient $\nabla_{\gamma} s^{2} \equiv \frac{\partial}{\partial u_{\gamma}} s^{2}$ if it is needed). 
 \item Find the correction $\Delta u^{(i)}_{\gamma}$ based on the direct iteration method described in eq.(\ref{fee:iteration2}) or eq.(\ref{fee:iteration3}), the conjugate gradient method or another local optimization methods, the parameters of CG potential for the next iteration is  $u^{(i+1)}_{\gamma} = u^{(i)}_{\gamma} + \Delta u^{(i)}_{\gamma} $. 
\item Do aforementioned two steps until $s^{2}$ is smaller than a preselected threshold, which could be determined by analyzing statistic error of $s^{2}$.
\end{enumerate}

\subsection{Pressure Correction} 
However, while the ensemble average values of any conformation function $B(x)$ in the AA model can be approximately obtained from $\langle B(x) \rangle_{cg}$ with a relative error not more than $s$, some interesting physical quantities, such as pressure, can not be written as the ensemble average of a common conformational function in the CG and AA models. Actually, the microscopic function of pressure   
\begin{eqnarray}
{\hat {\cal P}}(x) = \frac{N k_{B} T}{V} - \frac{1}{d V} L \frac{\partial W(x)}{\partial L},
\end{eqnarray} 
has explicit dependence on potential energy surface $W(x)$. Here $V=L^{d}$ is the volume of the $d$-dimensional simulation box with length $L$, and $W(x)$ is dependent on $L$ while scaling the real conformation $x$ as the dimensionless one $z=x/L$. 
For example, given the assumption that the force filed of CG model is pairwise central additive and $d=3$, the pressure is thus able to be expressed as an integral over RDF \cite{Hansen2006} and further can be discretized in terms of potential parameters, 
\begin{eqnarray}
  \label{fee:pc}
  {\hat {\cal P}}_{cg} (x) &=& \frac{2 \pi \rho N}{3V} \int_0^{\infty} [3 r^2 {\hat g}(r;x) + r^3 {\hat g}^{\prime}(r;x)] u(r) dr + \frac{\rho}{\beta} \nonumber \\
   &=& \Upsilon^{\gamma}(x) u_{\gamma} + \rho k_{B} T, 
\end{eqnarray}
where $\rho = N/V$ is the number density of CG particles, ${\hat g}(r;x)$ is value of RDF of special CG conformation $x$ at inter-particle distance $r$, ${\hat g}^{\prime}(r;x) \equiv \frac{\partial {\hat g}(r;x)}{\partial r}$, and 
\begin{eqnarray} 
\Upsilon^{\gamma}(x) \equiv \frac{2 \pi \rho N \Delta r}{3V}[3 r_{\gamma}^2 {\hat g}(r_{\gamma};x) + r_{\gamma}^3 {\hat g}^{\prime}(r_{\gamma};x)]. 
\end{eqnarray} 
Here we already suppose $r^{3} {\hat g}(r;x) u(r)= 0$ while $r=0$ and $r \rightarrow \infty$, which is usually true. 
It is clearly, ${\hat {\cal P}}_{cg}(x)$ is explicitly dependent on the parameters $u_{\gamma}$ of the CG force field, thus  
the macroscopic pressure, ${\cal P}_{cg}=\langle {\hat {\cal P}}_{cg}(x) \rangle_{cg}$, not only depends on the equilibrium distribution of CG, but also explicitly depends on the potential surface, $U(x;u_{\lambda})$, itself. 
Thus, the error of the macroscopic pressure, $\epsilon_{\cal P}=|\langle {\hat {\cal P}}_{aa}(q) \rangle_{aa}-\langle {\hat {\cal P}}_{cg}(x) \rangle_{cg}|$ is, 
\begin{eqnarray} 
 \epsilon_{\cal P} = |\langle {\hat {\cal P}}_{aa}-{\hat {\cal P}}_{cg} \rangle_{aa}  
+ \langle {\hat {\cal P}}_{cg} \rangle_{aa} - \langle P_{cg} \rangle_{cg}|. 
\label{eq:pressure-error}
\end{eqnarray} 
While the second term in the right side of eq.(\ref{eq:pressure-error}) is limited by the $s_{aa \rightarrow cg}$, the first term might be large since the microscopic pressure ${\hat {\cal P}}$ has different dependence on $x$ in CG model from that in AA system. 
For reconstructing the pressure in CG model, extra efforts should be taken. One simple way is to directly use a penalty function to define a pseudo distance, 
\begin{equation}
  \label{pc:s2p}
  s^2_P \equiv s^2_{aa \rightarrow cg} + \alpha [\langle {\hat {\cal P}}_{cg}(x) \rangle_{cg} - \langle {\hat {\cal P}}_{aa}(q) \rangle_{aa}]^2, 
\end{equation}
where 
$\alpha$ is a positive constant. 
The gradient of the pseudo distance now becomes 
\begin{equation}
  \label{pc:gradient}
  \frac{\partial s^{2}_{P}}{\partial u_{\gamma}} = \frac{\partial s^2_{aa \rightarrow cg}}{\partial u_{\gamma}} + 2 \alpha [\langle {\hat {\cal P}}_{cg} \rangle_{cg} - \langle {\hat {\cal P}}_{aa} \rangle_{aa} ] \frac{\partial}{\partial u_{\gamma}} \langle {\hat {\cal P}}_{cg} \rangle_{cg}, 
\end{equation}
where 
\begin{eqnarray}
\frac{\partial}{ \partial u_{\gamma} } \langle {\hat {\cal P} }_{cg} \rangle_{cg} = - \beta \langle \delta_{cg} f^{\gamma}(x) \; {\hat {\cal P}}_{cg}(x) \rangle_{cg} + \langle \Upsilon^{\gamma}(x) \rangle_{cg}. 
\end{eqnarray}
With eq.(\ref{pc:s2p}) and eq.(\ref{pc:gradient}), 
we can minimize $s^{2}_{P}$ by using usual local minimization methods, such as the conjugate gradient method. 
In eq.(\ref{pc:s2p}), the penalty coefficient $\alpha$ can be any positive number in principle. However, it is possible to make the pressure correction more consistent with the distribution matching scheme by defining the value of penalty coefficient in following way: First, recall the expansion expression given by eq.(\ref{eq:s2}), if $\delta_{aa}B^{\mu}$ are orthogonal to each other but not necessarily have been normalized, then eq.(\ref{eq:s2}) can be re-expressed as 
$s^2=\sum_{\mu} G_{\mu\mu} \langle \delta_{aa} B^{\mu}\rangle_{cg}^2 $, where $G_{\mu\mu}=\langle (\delta_{aa} B^{\mu})^{2}  \rangle_{aa}^{-1}$. 
Now if we add a hybrid basis function $\delta_{aa} {\hat {\cal P}}(x) \equiv {\hat {\cal P}}_{cg}(x) - \langle {\hat {\cal P}}_{aa} \rangle_{aa}$ in the basis function set and neglect the correlation between the function and the rest of basis functions, we are able to give an appropriate value for the penalty coefficient, that is the reverse variance of ${\hat {\cal P}}_{aa}$ in the AA system, {\it i.e.}, 
\begin{eqnarray}
\alpha = [\langle {\hat {\cal P}}_{aa}^{2}(q) \rangle_{aa} - \langle {\hat {\cal P}}_{aa}(q) \rangle_{aa}^{2}]^{-1}.
\end{eqnarray} 

It is possible to include $\delta_{aa} {\hat {\cal P}}(x)$ (and its some functions) in the set of basis functions, and expand the probability ratio in an extended space, thus also take into account the correlation of the new pressure-related basis functions with the other basis functions. 
 It might further improve the constructed CG model in reproducing overall character of original system. 
 However, for simplification, here we only use eq.(\ref{pc:s2p}) to illustrate the possibility of involving pressure matching in optimization of CG models, more works will be done in the future. 

\section{Test Case: One site water model} \label{sec:experiment}
 In this section, we use PMCG to construct an effective (CG) force field for one site water model. 
We choose the TIP3P water model\cite{jorgensen:926} as AA  system, which is widely applied in atomistic molecular simulations of water systems. 
 The illustration of applying PMCG to optimize effective CG force field is not dependent on the selection of higher-resolution model, for example, we could directly construct the one-site water from the first principle quantum mechanics based simulations of water, which is very expensive and usually only treats hundreds water molecules and is not easy to reach to a nanosecond. In comparison with the {\emph ab initio} water model, usual all-atomic water model, such as TIP3P, can reach tens of thousands of water molecules and reach hundreds of nanoseconds or longer. 
 The TIP3P water model treat water molecular as a rigid three sites model, and its force field involves only non-bonded interactions, {\it i.e.}, the electrostatic interactions and the Lennnard-Jones potential. 
The TIP3P simulation of water was carried out in the constant NVT condition, where the number of water molecules $N= 216$ in a cubic simulation box with the volume $6.4585 \; {\mathrm nm^3}$ (corresponding the density of water as $1.0 \; {\mathrm g \cdot cm^{-3} }$) at the temperature $T=300\; {\mathrm K}$. In one site CG model, the water molecules are replaced by a spherically symmetrical site mapped by the center of mass of water molecules with the mass $2 m_{H} + m_{O}$. Both the CG model and the TIP3P model are simulated by the MD package NAMD2.5~\cite{namd} with our modification for coarse-graining MD simulations. In all simulations, the Lennnard-Jones potentials are gradually switched off from radius $r_{switch}=0.8 \; {\mathrm nm}$ to $r_{cutoff}=0.9 \; {\mathrm nm}$, 
the electrostatic force is calculated based on particle mesh Ewald (PME) method. 
Langevin thermostat with the a damping coefficient of $5.0 \; {\mathrm ps^{-1}}$ is used; a length $20 {\mathrm ns}$  trajectory was generated from TIP3P simulation, and conformations are collected in a frequency of every $0.4 \; {\mathrm ps}$ after a segment of equilibrating  simulations. 

The calculation of electrostatic  interactions in TIP3P (or another atomistic) water model is the bottleneck of simulations. 
CG water model without considering the atomic details but treating each molecule as one particle and using effective pair-additive interaction to mimic electrostatic potential is usually accelerate simulation about one magnitude of order at least.
Here we construct one of the kind CG model to illustrate our general CG method. In the CG model, 
a tabular interaction $\{u_{\gamma}\}$ at some pair distances $\{ r_{\gamma}, \gamma=1, \cdots,\}$ is used to intrapolate the interaction value $u(r)$ of a pair of CG particle at any distance $r$. Thus the tabular interaction can be very various to mimic the effective potential of electrostatic interaction and the Lennard-Jones potentials between oxygen atoms, but does not increase any computation cost in comparison with any analytic function as the effective interaction. 
 For optimizing the parameters $\{u_{\gamma}\}$, we literately run many short segment of CG simulations, in the same condition with that of the TIP3P simulation. The length of each segment of trajectory is $800 \; {\mathrm ps}$, with a sampling interval of $0.2 \; {\mathrm ps}$ after the first $10\; \mathrm{ps}$ trajectory for equilibrating. After each $800 \; {\mathrm ps}$ segment, we update the parameters $\{u_{\gamma}\}$, repeat the process until the distance of CG model from the AA model, $s^{2}$, be minimal. 
 
In the paper, we only apply the microscopic RDF ${\hat g}(r; x)$ at different particle-pair distance $r$ 
 as the basis functions to illustrate the method. It is directly to use more basis functions, such as multiple-body correlations, local orientational orders to obtain more robust CG force field. All these preselected physical quantities are applied to estimate $s^{2}$ with the inverse correlation matrix $G_{\mu\nu}$. 
 As we already mentioned in the previous section, ${\hat g}(r; x)$ is the probability to find two particles (water molecules here) with distance $r$ in the particular conformation $x$, {\emph i.e.}, ${\hat g}(r; x)$ is the microscopic correspondence of the usual RDF $g(r)$, which characterizes the structure of liquid, and is expected to be sufficient in estimate $s^2$. 
In the practical application, we divide the particle-pair distance $r$ into some small bins, and use the value of ${\hat g}(r; x)$ inside each bin 
as basis function, ${\hat g}_{\mu}(x) = {\hat g}(r_{\mu}; x), \mu=0, \cdots, m=99$. Here $r_{\mu} = r_{min} + \mu \Delta r$,  
 with $r_{min}=0.1\; \mathrm{nm}$, and $\Delta r=0.008\; {\mathrm nm}$. 
 Without loss any generality, these values of microscopic RDF are first scaled then as basis functions, {\emph i.e.}, 
\begin{equation}
  \label{eq:basis}
  \delta_{aa} B^{\mu}(x) = \frac{{\hat g}_{\mu}(x) - \langle {\hat g}_{\mu}(x) \rangle_{aa} }{\sigma_{aa}^{\mu} + \varepsilon_A}
\end{equation}
where the scaling factor $\sigma_{aa}^{\mu} = \sqrt{ \langle {\hat g}^2_{\mu}(x)\rangle_{aa}-\langle {\hat g}_{\mu}(x)\rangle^2_{aa}}$, the fluctuation of ${\hat g}_{\mu}(x)$ in the AA  model, is applied to make value of every basis function be in $O(1)$ for getting better numerical stability in calculating the inverse matrix $G_{\mu\nu}$. 
$\varepsilon_A$ is a small positive value that makes sure the denominator is nonzero. 
 
\begin{figure}[h]
  \centering
  \includegraphics[width=0.5\textwidth]{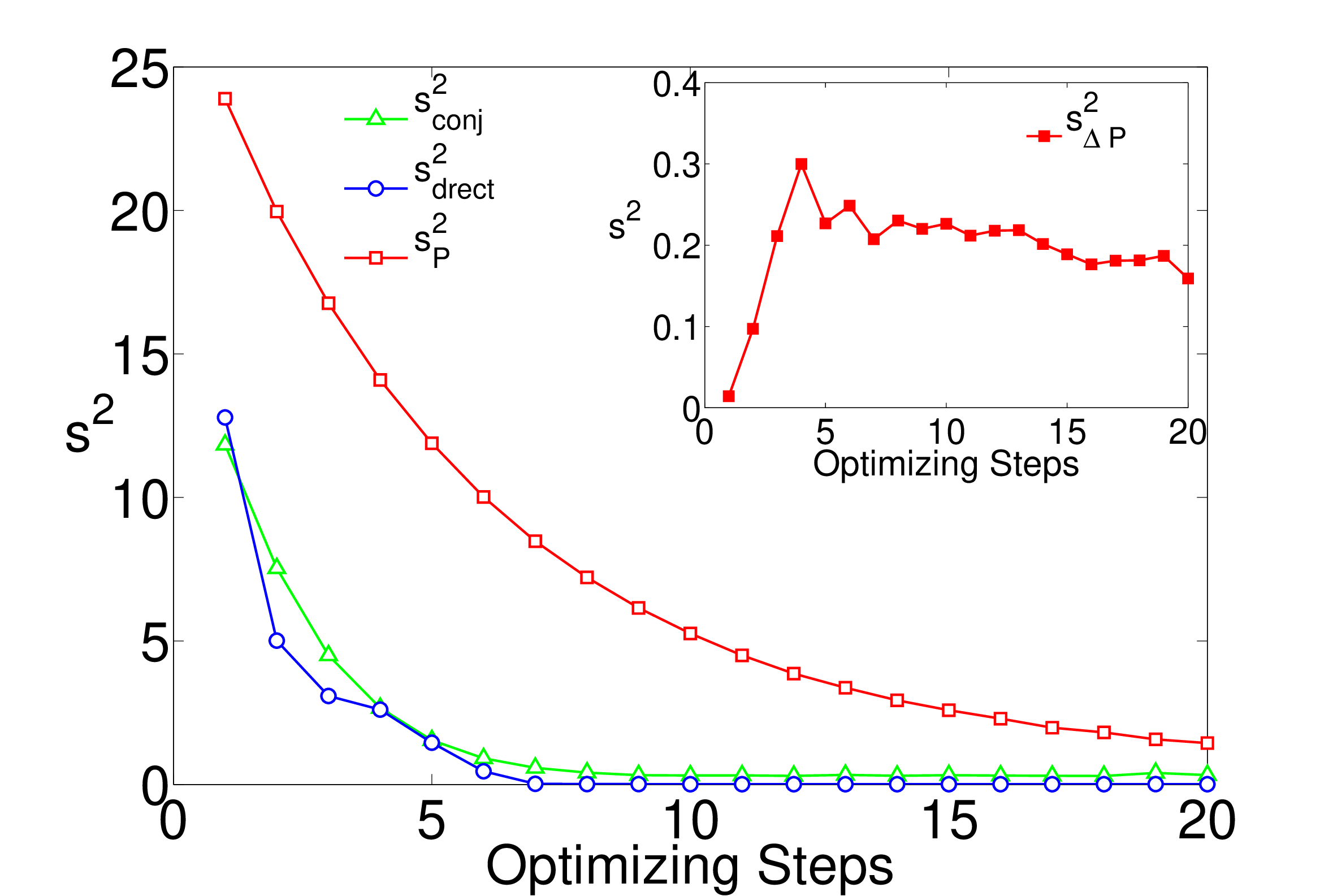}
  \caption{Evolution of $s^2$ (without pressure correction) and $s^2_P$ (with pressure correction) in the first $20$ iteration steps using different optimizing algorithm. 
  The subscript notion \textit{conj} refers to the conjugate gradient method; \textit{direct} means the direct iteration method (see text). 
 At the $20$th step, $s^2(20)$ is $0.3$ with the conjugate gradient method, is $0.007$ with the direct iteration method, 
 meanwhile $s^2_{P}(20)$ is $1.4$ and $s^2_{\Delta P} = 0.16$
}
  \label{fig:s2}
\end{figure}

\begin{figure}[h]
  \centering
  \includegraphics[width=0.5\textwidth]{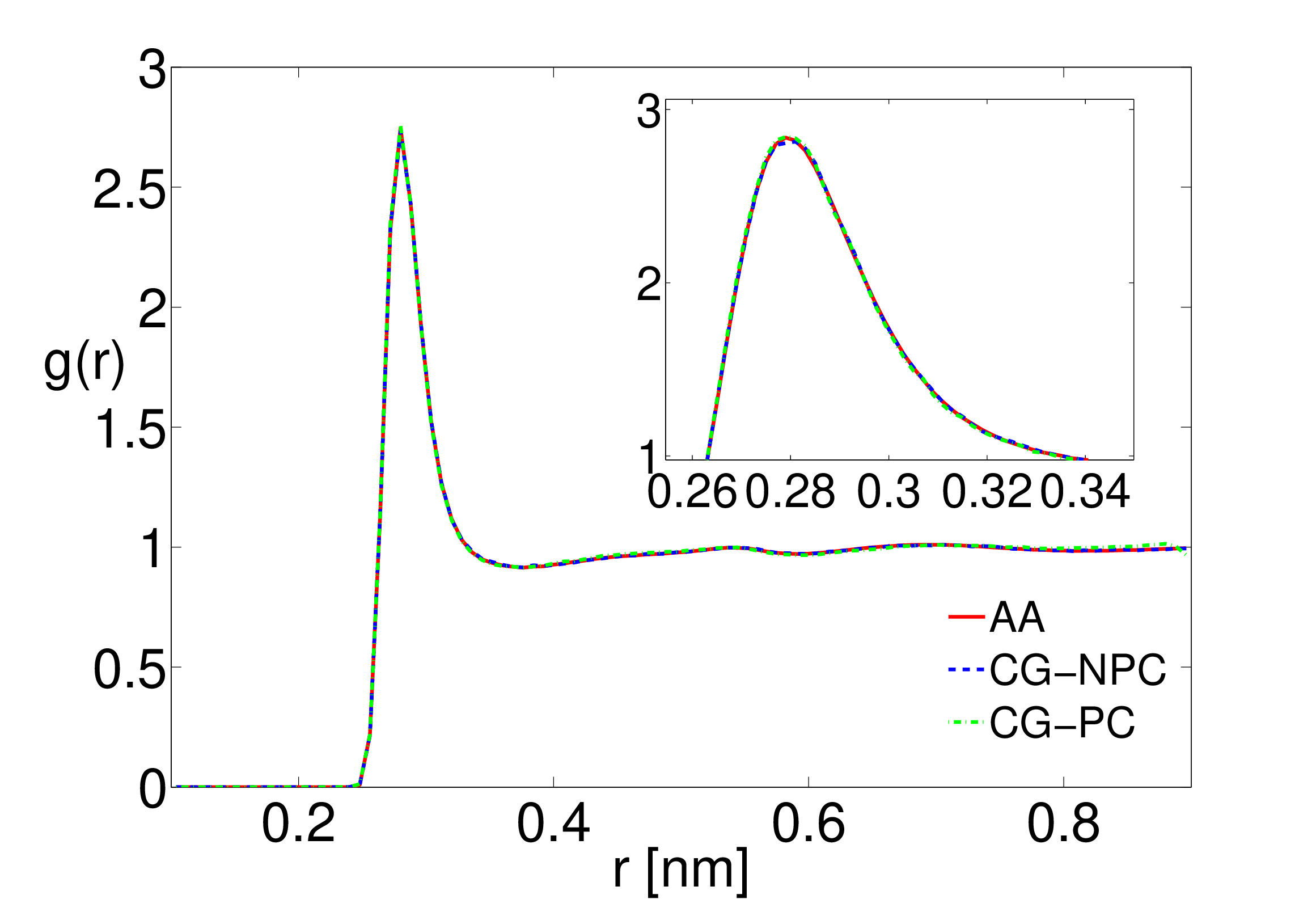}
  \caption{Comparison between center of mass RDF of TIP3P water model and site-site RDF of CG water model. Both simulations were performed under the constant NVT ensemble at temperature of $300\;{\mathrm{K}}$ and density of $1.0 \; {\mathrm{g/cm^3}}$. The inset shows an enlarged region of the first peak. The $s^2$ between the CG model without the pressure correction (CG-NPC) and the TIP3P model is $0.001$, and $s^2_P$ between pressure corrected CG model (CG-PC) and TIP3P model is about $0.084$.}
  \label{fig:gofr}
\end{figure}

We starts from potential function defined in eq.(\ref{eq:initialGuess}) and minimize the $s^2$ with both the conjugate gradient method and the direct iteration method described by eq.(\ref{fee:iteration2}). Both of the methods work effectively except for the fact that the direct iteration method minimizes $s^2$ slightly faster than the conjugate gradient method in the case. As demonstrates in Fig.~\ref{fig:s2}, when using the direct iteration method $s^2$ decrease from $12.0$ to $0.01$ within the first $10$ iteration steps, on the other hand, the conjugate gradient method need more than $15$ iteration steps to decrease $s^2$ from $11.9$ to $0.3$. 
 When the distance is smaller than $0.001$, RDF, which we used to construct the basis functions, of referenced TIP3P model can be reproduce by CG model remarkable accurately, as demonstrated in Fig.~\ref{fig:gofr}. 

\begin{figure}[h]
  \centering
  \includegraphics[width=0.5\textwidth]{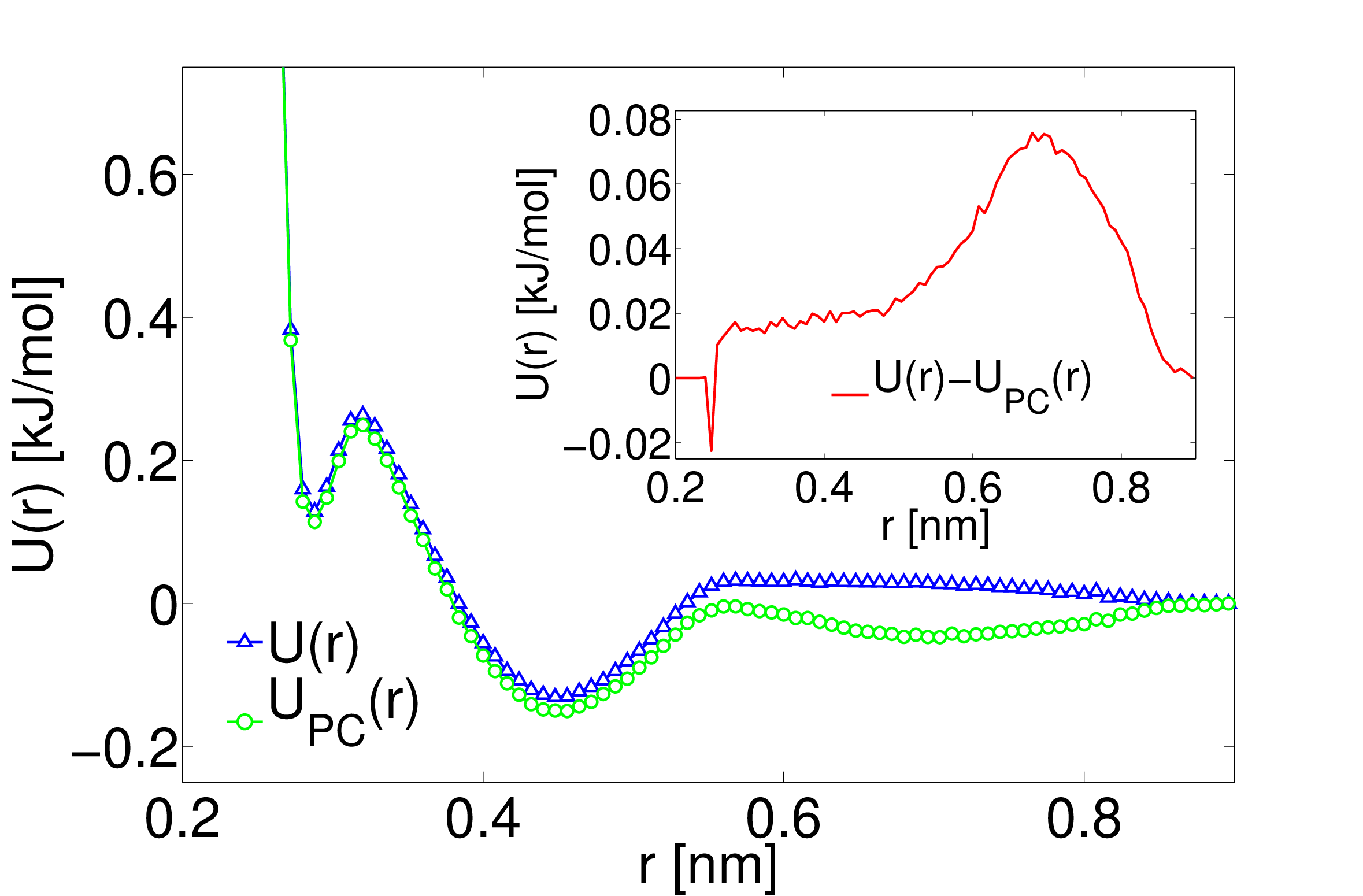}
  \caption{The effective pair potential for CG with and without pressure correction; the inset shows the difference between the two. $s^2$ and $s^2_P$ of the two potentials are the same with those of Fig. \ref{fig:gofr}.}
  \label{fig:cgp}
\end{figure}

The pressure in CG model can be very different from that in AA system if only RDF is fitted. 
Former researchers~\cite{ib:jcc2002,wang2009} used to correct CG force field by adding a linear item to the effective pair potential to reproduce pressure of AA system in CG model. 
We declare that the linear correction is not the best compromise to revise pressure while having reasonably small effect on the accuracy of structure properties' reconstruction. The effective pair potential of CG before and after the pressure correction described by eq.(\ref{pc:s2p}) was shown in Fig.~\ref{fig:cgp}. Correspondingly, the evolution of $s^2_P$ is shown in Fig.~\ref{fig:s2};  RDF after the pressure correction is also shown in Fig.~\ref{fig:gofr} as comparison. Fig.~\ref{fig:cgp} clearly  demonstrates that the pressure correction has change the pair potential more in the large $r$ region, {\it i.e.}, $0.6 \sim 0.7 \; {\mathrm nm}$, than in the small $r$ region, {\it i.e.}, $0.2 \sim 0.5 \; {\mathrm nm}$. 
 
\begin{figure}[h]
  \centering
  \includegraphics[width=0.5\textwidth]{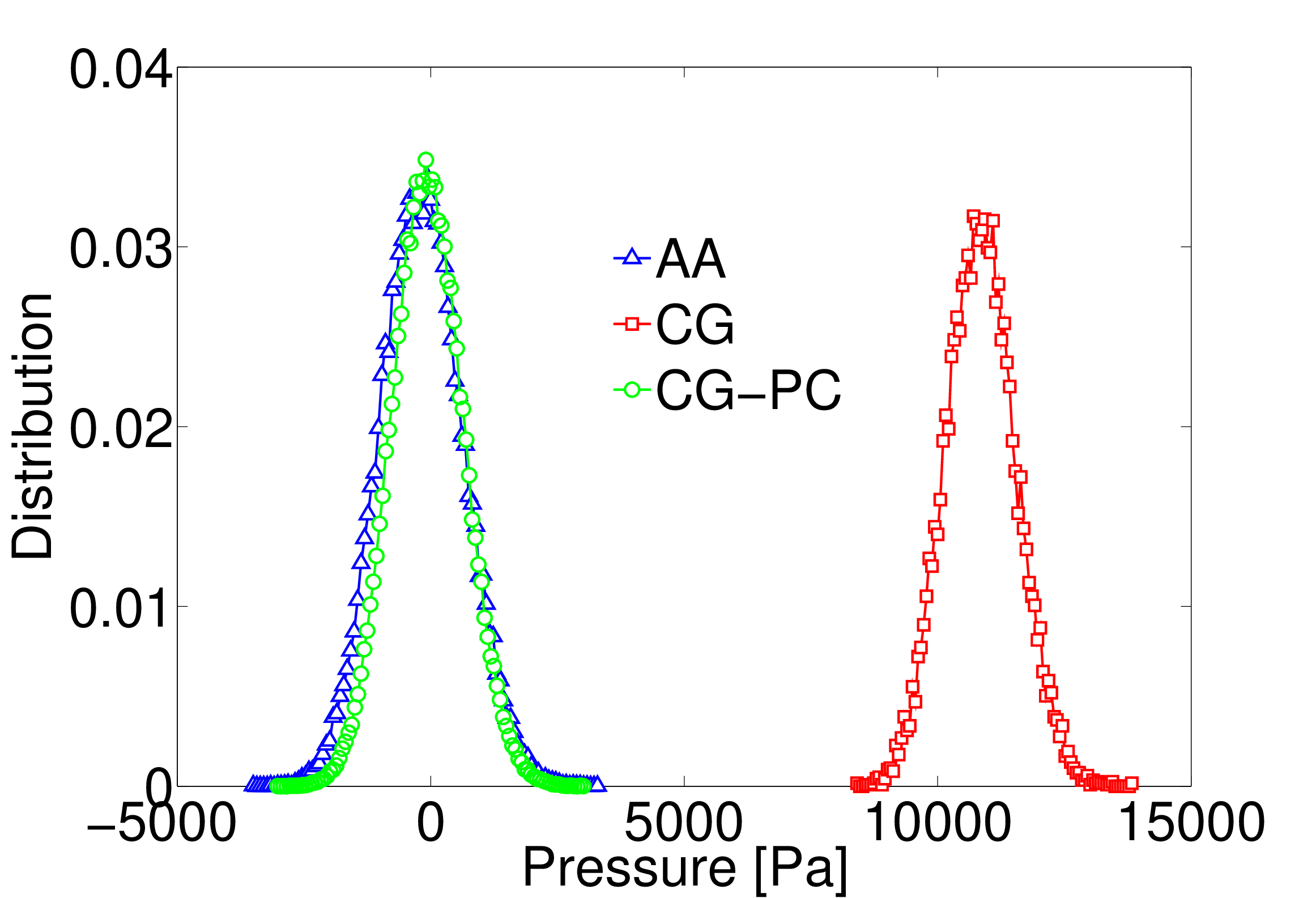}
  \caption{The pressure distribution in all-atom model (triangle line), coarse-grained models with pressure correction (circle line) and without pressure correction (square line). The average pressure of these distributions are, accordingly, $-184.3 \; {\mathrm{bar}}$, $-64.5 \; {\mathrm{bar}}$ and $1.08 \times 10^4 \; {\mathrm{bar}}$, resepctively.}
  \label{fig:Ps}
\end{figure}

As for the one site water model, as shown in Fig.~\ref{fig:Ps}, the variance of pressure in the AA system, $\sigma(\delta_{aa} {\cal P}) = 818.1 {\mathrm bar}$, which is equal to 
$0.012 \; {\mathrm kcal \cdot mol^{-1} \cdot \AA^{-3}}$, we use 
$1/\sigma^2(\delta_{aa} {\cal P}) \approx 7200 \; {\mathrm mol^{2} \cdot \AA^{6} \cdot kcal^{-2}}$ as the value for $\alpha$ in the pressure correction. We carried on the pressure correction after $s^2$ had decreased to $0.014$. The evolution of $s^2_P$ is presented in Fig.~\ref{fig:s2}. It is worth noticing that in the first four steps, minimization of $s^2_{\Delta P}$ makes concession to that of $s^2_P$ and increase from $0.014$ to $0.3$ due to the fact that pressure difference is too large. However, from the fifth step on,
both of them can decrease smoothly. This is largely because pressure difference is numerically independent on RDF differences. Correction to the potential at this moment is roughly divided into two parts, one is the fine tune which add short wave length modification function to pair interaction; the other part is the global shift which change the potential function significantly while holding the fine structure.

\begin{figure}[h]
  \centering
  \includegraphics[width=0.5\textwidth]{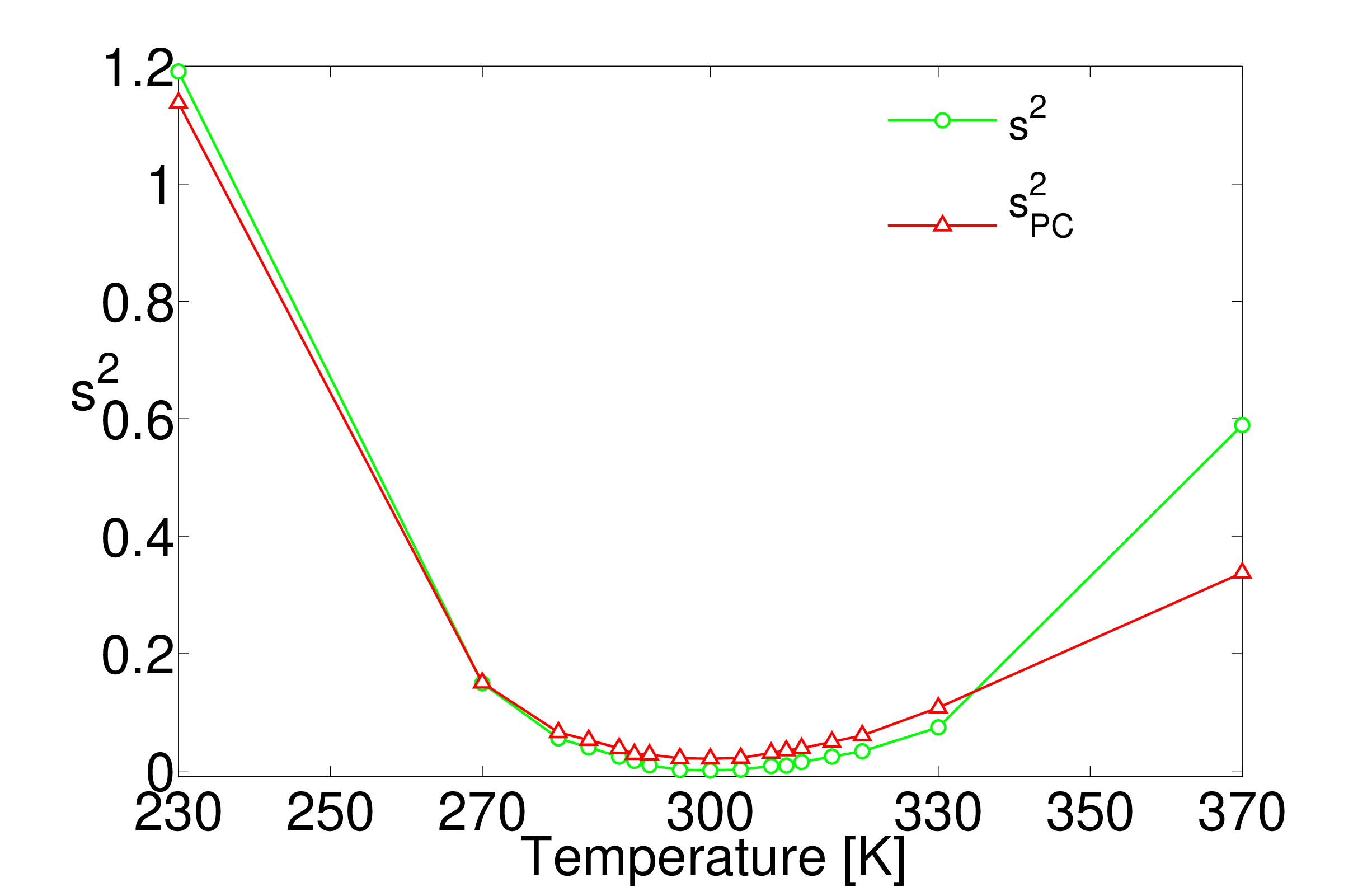}
  \caption{Comparison of extensibility of pair potentials on different temperatures. The initial potentials is obtained in $T=300 \; {\mathrm{K}}$ from different methods, using versus not using pressure correction in particular, and then is tested in other temperatures from $230\; {\mathrm{K}}$ to $370\;{\mathrm{K}}$. The results are then compared to TIP3P simulations at the same temperature so as to calculate the free energy distances ($s^2$). The circle line shows the dependence of $s^2$ on temperature for non-pressure correction pair potential, its value at $T=300\;{\mathrm{K}}$ is $0.001$; the triangle line show the results of the potential with pressure correction, its minimum value is $0.026$.}
  \label{fig:s2ext}
\end{figure}

One of the well known dilemmas in CG techniques is that the pressure correction would lead to less accuracy in representing of other properties\cite{wang2009, johnson:144509,Louis:2002,hoef:1520} such as isothermal compressibility and conformance of RDF. This effect is shown in Fig.~(\ref{fig:s2ext}): when pressure consistency is taken into account, $s^2$ increases from $0.001$ to the value of $0.026$. However on the other hand, CG with pressure correction shows greater extensibility than that without pressure correction at temperature $T=370 \; {\mathrm K}$ and $T=230\;{\mathrm K}$. 
It indicates the additional pressure correction makes the CG model be more consistent with the AA system. 
 Another appealing feature Fig.~(\ref{fig:s2ext}) shows is that in the temperature region of $T_{eff} =\{T|  290 \leq T \leq 312 \} $ , $s^2_{P} \leq 0.04$ and $s^2 \leq 0.017 $, both of which are very small value. Especially that in $T_{eff}$, $s^2$ is smaller that the best case of $s^2_{P}$. Since the RDF corresponding to $s^2_{P} = 0.026$ is almost indecipherable, we can safely declare that $T_{eff}$ is the effective temperature region for the potential obtained from our method.
 
\section{CONCLUSION}
The present work introduces a new methodology, the distribution matching method, to optimize CG force field effectively and efficiently. Consistency condition between CG and detailed atomistic model is given and reinterpreted as a requirement of matching of distributions thus equilibrium average of a set of sufficient and independent basis functions through the distribution expansion analysis. Based on the analysis, we proposed a two steps protocol to construct effective CG force filed. The first step is to expand phase spaces differences between CG and AA model as linear combination of basis functions. In this part orthogonalization technique is suggested to avoid singularity led by linear dependencies among basis functions. The second step is to minimize the defined difference in whole the conformational space. Two different minimization approaches are introduced in this part, namely the conjugate gradient approach and the direct iteration approach. Both two approaches are demonstrated in the case of being applied to fit pairwise additive force field. With aforementioned two steps protocol we have been fully able to construct effective pair additive force field for CG models. We test this statement by applying the formulas to building one site model for TIP3P water model. 
 Considering that pressure consistency is required in some situations we propose a method to correct pressure in accordance previous two steps protocol. It is enlightened by constraint optimization techniques. By adding a penalty item to the definition of phase space difference we are able to limit the pressure deviance. The effectiveness of pressure correction method is also verified in one-site CG water model. 
 The present methodology is encouraging, its capability to optimize CG models indicates a wide applications in multiscale simulations.

\acknowledgements{
The work is supported by NSFC under Grant No. 11175250. The authors thank Prof. Bo Zheng for useful discussions. X.Z thanks the financial supports of the Hundred of Talents Program in Chinese Academy of Sciences. The authors also thank the support of Max Planck Society and the Korea Ministry of Education, Science and Technology, part of the work had been finished while they worked at Asia Pacific Center for Theoretical Physics under the support.  
}


\

\newpage

\end{document}